\documentstyle[aps,prl,multicol,epsf]{revtex}
\title{\bf Aging in a Model of Self-Organized Criticality}

\author{Stefan Boettcher$^{1,2,3}$ and Maya Paczuski$^4$}
\address{
$^1$Center for Theoretical Studies of Physical Systems, Clark Atlanta
University, Atlanta, GA 30314\\
$^2$Department of Physics and Astronomy, The
University of Oklahoma, Norman, OK 73019-0225\\
$^3$Center for Nonlinear Studies,
MS-B258, Los Alamos National Laboratory, Los Alamos, NM 87545\\
$^4$Department of Physics, University of Houston, Houston TX 77204-5506}
\date{\today}
\begin{document}
\maketitle

\begin{abstract}

Temporal autocorrelation functions for avalanches in the Bak-Sneppen
model display aging behavior similar to glassy systems.  Numerical
simulations show that they decay as power laws with two distinct
regimes separated by a time scale which is the waiting time, or age,
of the avalanche.  Thus, time-translational invariance is dynamically
broken.  The critical exponent of the initial decay is that of the
familiar stationary dynamics while a new critical exponent for the
late-time behavior appears.  This new exponent characterizes a
non-stationary regime that has not been previously considered in the
context of self-organized criticality.\break 
PACS numbers: 64.60.Lx, 05.40.+j, 05.70.Ln
\end{abstract}

\begin{multicols}{2}
Self-organized criticality (SOC) \cite{BTW} describes a general
property of slowly-driven dissipative systems with many degrees of
freedom to reach a stationary state where they evolve
intermittently in terms of bursts spanning all
scales up to the system size.  Many natural avalanche-like phenomena
have been represented using this idea, including earthquakes
extinction events in biological evolution,
and landscape formation \cite{per}.
Recently, SOC has been observed in controlled laboratory experiments
on rice piles \cite{Frette}.  Theoretical models of rice piles
\cite{Oslo} are related to a variety of different physical systems by
universality \cite{PaBo}.  The emergence of long-term memory in the
self-organized critical state has been demonstrated analytically
 for a multi-trait evolution model \cite{BoPa2}, a
variant of the Bak-Sneppen model \cite{B+S}.

Here we show that the self-organized critical state in the Bak-Sneppen
model exhibits aging behavior that is reminiscent of glassy systems
\cite{slowdyn}.  We numerically measure two-time autocorrelation
functions for avalanches and observe a power-law decay with two
distinct regimes.  The early time power law regime is that of the
familiar stationary dynamics.  The late time regime has a new critical
exponent characterizing the history-dependent relaxation behavior of
the model.  The time scale separating the early and late time regimes
is the age of the avalanche.  Rescaling the autocorrelations measured
in units of the respective age collapses the data onto a scaling
function of a single variable.  The distinct late time regime signals
a breaking of time-translational invariance arising from the system's
ability to store information from past events over arbitrarily long
times.

``Aging'' is an important phenomenon observed experimentally in glassy
materials where relaxation behavior depends on the previous history of
the system \cite{AHN}.  For instance, consider a spin glass quenched
below the glass transition at time $t=0$ in the presence of a small
magnetic field.  Throughout the sample, domains of various pure states
develop that grow with characteristic size $R(t)$.  At time $t=t_w$,
the magnetic field is turned off and the system's response in the form
of its remanent magnetization is measured.  Initially, the response
function is only sensitive to the pure, quasi-equilibrium states in
their distinct domains.  But after an additional time scale related to
the waiting time, $t_w$, the response spans entire domains and slows
down when it experiences the non-equilibrium state the sample
possesses as a whole.  A
simple scaling form holds for some glassy systems
 \cite{scaling_glasses}.  The
autocorrelation function for the remanent magnetization is given by
\begin{equation}
C(t+t_w,t_w) = t^{-\mu}f(t/t_w) \quad ,
\label{scaleq}
\end{equation}
where $f(x\rightarrow 0)\rightarrow constant$ and $f(x>>1)\sim x^{-r}$
with $r >0$.  Similar aging behavior also
appears in a variety of other problems such as coarsening dynamics,
directed polymers in random media, etc. \cite{other}.

Generally, this type of
 correlation function can be measured for avalanches in the
self-organized critical state (which will be specified below for the
Bak-Sneppen model).  Here, we define the waiting time,
$t_w$, to be the time since the beginning of the avalanche.  In order
to eliminate trivial effects of norm conservation \cite{trivial}
referring to the fact that avalanches die, we only consider the
dynamics of infinite avalanches, or those that survive longer than any
time scale we consider.
We focus on a simple quantity, $P_{\rm first}(t)$, measuring
the first returns of the activity to a given site.  A power law
distribution for $P_{\rm first}(t)$ has been measured numerically for
a variety of different SOC models \cite{scaling} by recording first
returns to every site visited during the avalanche.  Here we measure
the probability $P_{\rm first}(t_w; t)$ for the infinite avalanche to
return after $t$ time steps to a site that was visited most recently
at time $t_w$ from the beginning of the avalanche.  Thus, to determine
the first-return probability, we take the age of the avalanche, $t_w$,
into account. While in the stationary state of SOC models the first
return distribution is generally $P_{\rm first}( t)\sim t^{-\tau_{\rm
 first}},~( t\to\infty)$, we find for the Bak-Sneppen model
\begin{eqnarray}
P_{\rm first}(t_w; t)\sim  t^{-\tau_{\rm first}} &
 f\left({ t \over t_w}\right)&\cr
\noalign{\medskip}
&f(x)&\sim\cases{ {\rm constant} &$(x\ll 1)$,\cr
\noalign{\medskip}
x^{-r} &$(x\gg 1)$.}
\label{pfirst}
\end{eqnarray}
The exponent $\tau_{\rm first}$ can be related to other critical
exponents via scaling relations for SOC \cite{scaling}.  The origin of
the coefficient $r$ is non-trivial, signaling the dynamical
breaking of time-translational invariance in the
infinite avalanche.  (For instance, the first-return probability to
any site $n$ for a random walker that started at the origin at time
$t'=0$ and passed site $n$ at time $t_w$ is clearly independent of
$t_w$ unless time translational invariance is explicitly broken.) In glassy
systems, the coefficient $r$ usually refers to off-equilibrium properties of
the system \cite{slowdyn}. The question then arises whether the
coefficient $r$ in Eq.~(\ref{pfirst}) can be related to the known
universal coefficients of the stationary SOC process, or whether it
describes new physics in the avalanche dynamics of the Bak-Sneppen model,
and possibly other models of SOC.

The Bak-Sneppen model \cite{B+S} has been studied intensely and with great
numerical accuracy in recent years. We refer to Ref.~\cite{scaling} for
a review of its many features and simply utilize those facts here.  The model
consists of random numbers $\lambda_i$ between 0 and 1, each occupying
a site $i$ on a lattice in, say, $d=1$. At each update step, the
smallest random number $\lambda_{min}(t)$ is located.  That site as
well as its two nearest neighbors each get new random numbers drawn
independently from a flat distribution between zero and one.  The
system evolves to a SOC state where almost all numbers have values
above $\lambda_{\rm c}$, with $\lambda_{\rm c}$ avalanches formed by
the remaining numbers below.  In our simulations, we have used the
equivalent branching process \cite{PMB} with $\lambda_{\rm c}=0.66702$
to eliminate any finite-size effects.  Initially, at time $t_a=0$, the
smallest threshold value is set equal to $\lambda_{\rm c}$ to start a
$\lambda_{\rm c}$ avalanche. In every update $t_a\to t_a+1$, only the
signal $\lambda_{\rm min}(t_a)$ and its two nearest neighbors receive
new threshold values.  At any time, we store only those threshold
values $\lambda_i<\lambda_{\rm c}$ that are part of the avalanche
because only those numbers can contribute to the signal, i. e. can ever
become smallest number. In addition, we keep a dynamic list of every
site that has ever held the signal at some time to determine the
first-return probabilities. It has been shown that
the Bak-Sneppen model in $d=1,2$ is a fractal
renewal process \cite{scaling},
i. e. activity is recurrent for each site  unless the avalanche
dies. Avalanches end when there are no
$\lambda_i<\lambda_{\rm c}$, and a new (independent) avalanche is
initiated with $t_a=0$.   If the avalanche survived longer than
$t_{co}=2^{27}$ then we store the data for the first returns; otherwise
the data is discarded.  This way, we probe properties of
infinite avalanches up to the cut-off time scale $t_{co}$.

\begin{figure}
\epsfxsize=2.2truein
\hskip 0.15truein\epsffile{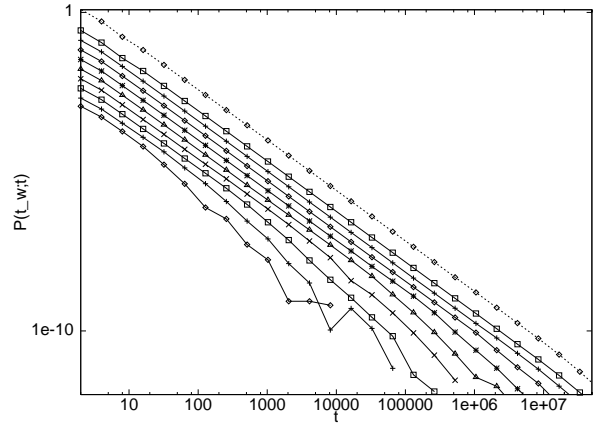}
\caption{\protect\label{bsd1fr}
\narrowtext
Log-log plot of the first-return probabilities $P_{\rm first}(t_w; t)$
as a function of $ t$ for various $t_w$ in the one
dimensional Bak-Sneppen model.
The graphs are consecutively labeled from bottom to
top by $i=1,~2,\ldots,~9$ where each graph contains data for all
$8^{i-1}\leq t_w\leq 8^i-1$.
To avoid overlaps, each distribution is
offset by a factor of $2^i$.
Initially, each graph falls like a power law
with the familiar coefficient $\tau_{\rm first}=1.58$. But for values
of $ t\gg t_w$, each graph crosses over into a new power law
regime with a larger coefficient. For reference, we have also plotted
the first return distribution for the combined data as a dashed line on
the right. Note the smooth power law behavior over six decades until
cut-off effects set in at about $10^7\approx 0.1~t_{\rm co}$.
}
\end{figure}

\begin{figure}
\epsfxsize=2.2truein
\hskip 0.15truein\epsffile{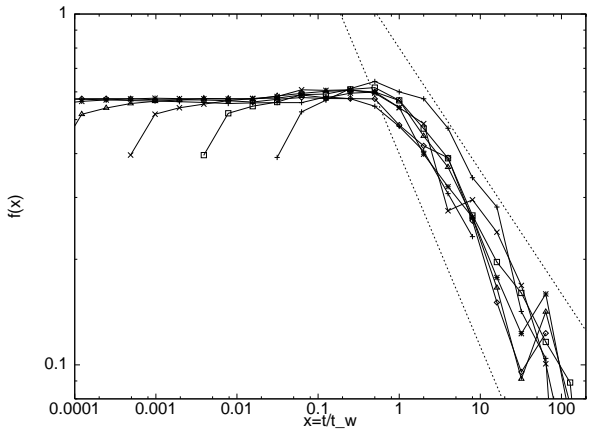}
\caption{\protect\label{bsd1scal}
Data collapse according to Eq. (\protect\ref{c1}) for the data in Fig.
\protect\ref{bsd1fr}
(using the same symbols), discarding data for $i=1$ and $i=9$ to eliminate
short-time and cut-off effects, respectively. The scaling function $f(x)$
is constant for small
argument and appears to follow a power law for $x\gg 1$ for about two
decades before the data gets too noisy. The two dashed lines that
bracket the power law tail correspond to $x^{-0.35}$ and $x^{-0.55}$.
We estimate $r=0.45\pm 0.10$.}
\end{figure}

At each update $t_a$ we determine the previous
time $t_w$ when the signal was on the
same site most recently (if ever). Then its first-return time is give by
$ t=t_a-t_w$, and we bin histograms labeled by
$i=\lceil{1\over 3}\log_2 t_w \rceil$ and $j=\lceil\log_2 t\rceil$.
The data is binned
logarithmically so that in each bin a comparable number of data points
is averaged over: for each increment of $j$, the width of the bins for
$t$ increases by a factor of $2$, while for each increment of $i$
the $t_w$ bins increase by a factor of $8$.
$P_{\rm first}(t_w;
t)$ for various values of $t_w$, and $P_{\rm first}( t)$ obtained
from the combined data, are plotted in Fig.~\ref{bsd1fr}.
Each graph refers to a different range for $t_w$, increasing by a factor
of $8$ each time from left to right. Each graph possesses two
distinct power law regimes, separated by a crossover.
To determine the form of the scaling function $f(x)$ for these graphs
according to Eqs.~(\ref{pfirst}), we note that the crossover appears to
scale linearly with $t_w$. Thus, we plot
\begin{eqnarray}
f(x)\sim  t^{\tau_{\rm first}} P_{\rm first}(t_w;t)
\quad {\rm with}\quad x={ t\over t_w}
\label{c1}
\end{eqnarray}
using $\tau_{\rm first}=1.58$ \cite{scaling}, see Fig.~\ref{bsd1scal}.  The data
collapses reasonably well onto a single curve, $f(x)$, which is
constant for small argument, and appears to fall like a power law with
coefficient $r=0.45\pm 0.10$ over two decades. The numerical
value for $r$ is difficult to extract with greater accuracy because it
refers to properties deep within the tail of the distribution, but it
is sufficiently accurate to prove the existence of a distinctly new
regime at time scales beyond $t_w$, indicating aging behavior.  The
data excludes the possibility that the effect is due to some hidden
cut-off (an exponential would have to be extremely weak to fall for
less than a decade over two decades in its argument), or due to mere
statistical noise.  Similar results were obtained for the two
dimensional Bak-Sneppen model where we find $r=0.25\pm 0.10$. Although
a smaller value of $\tau_{\rm first}=1.28$ in the two dimensional
Bak-Sneppen model leads to many more events in the tail of the
autocorrelation functions, our numerical results in that case are less
conclusive due to apparent significant corrections to scaling in the
distribution for the first returns \cite{else}.

  As written, the scaling theory of Ref.  \cite{scaling} does not
  describe the origin of the coefficient $r$ since it considers only
  stationary properties of avalanches where the
  breaking of time translational invariance
  does not appear.  We have considered domain
  growth approaches to explain the origin of aging, without success.
  For instance, the activity in an evolving avalanche
  spreads over a domain that is growing with a characteristic length
  scale $R(t)\sim t^{1/D}$ \cite{scaling}.  Thus, between $t_w$ and
  $t_w+t$, the domain covered by the avalanche has grown and might
  have decreased the probability to return to a particular site by a
  fraction $(1+t/t_w)^{-d/D}$, implying $r=d/D$.  For $d=1$, $D\approx
  2.43$ \cite{scaling}, which agrees well with the result for $r$, but
  for $d=2$, $D\approx 2.92$, which is inconsistent with the above
  result for $r$.  However only few sites inside an avalanche domain
  are actually part of the avalanche at any one time, their number
  merely growing like $t^{d_s}$ with $d_s\approx 0.11$ and $0.25$ in
  $d=1$ and $2$ \cite{scaling}.  Assuming that $r=d_s$ is inconsistent
  with the measured value of $r$ for $d=1$.  While similar geometrical
  arguments may
  quantitatively explain observed aging behavior in systems
  whose dynamics is dominated by the coarsening of domains
  \cite{Mezard}, it appears that aging in the Bak-Sneppen model and
  possibly other SOC models is more intricately linked to the
  long term memory of the process which can distinguish sites
  that were last visited early in the avalanche from those visited much
  later.

We have also investigated the possibility of aging in other SOC models
such as the Abelian sandpile due to Bak, Tang, and Wiesenfeld (BTW)
\cite{BTW} and the Manna two-state sandpile model \cite{Manna}, which
will be reported in detail elsewhere \cite{else}.  In short, numerical
measurements of the first return probabilities for the Manna model
show no sign of any age dependence at all.  The BTW model \cite{BTW}
does show nontrivial aging behavior which is quite different from that
of the Bak-Sneppen model,
i.e.  neither its scaling variable nor its scaling function is as
simple as Eq.~(\ref{pfirst}).  The appearance of aging in the BTW
model demonstrates at least a weak memory not included in the
discussion
of Ref. \cite{Marsili}.  Despite the gross similarities between
the BTW and the Manna model, our results suggest that the avalanche
dynamics in both models is completely different; in fact they have
been found to belong to different universality classes \cite{Biham}.
We do not know yet to what extent aging  is a
general property of self-organized critical systems.

It is well known from the study of amorphous materials that systems
with hierarchically ordered dynamics, where slow degrees of freedom
can only advance after faster ones have moved collectively, typically
lead to slow relaxation behavior \cite{PSAA}, similar to the power-law
distributed avalanches in SOC.  {}For instance, frustration has
been identified as a cause for slow relaxation in
spin systems, requiring cooperative behavior between ever more distant
spins to further decrease their collective energy.
Phenomenological models of relaxation in glassy systems
\cite{Bouchaud} attribute slow relaxation and aging
to the complicated phase-space structure, with a highly degenerate
equilibrium state, and with a hierarchy of ever-higher barriers slowing the
approach to any state of lower energy, that originates from the
inherent randomness in the microscopic interactions.  Previously,
explicitly hierarchical models of diffusion on ultrametric trees have
been invoked to describe experimental findings for such glassy systems
\cite{O+S}, including their aging behavior \cite{H+S}.

Remarkably, the Bak-Sneppen mechanism shares many of these ``glassy''
features. The far-from-equilibrium dynamics is characterized by a large
pool of virtually inert (``frozen'') threshold values, equally distributed
above $\lambda_{\rm c}$, assuming one among an infinite degeneracy of
quasi-stable configurations that the process is attracted to. The update process
perturbs the system, inducing avalanches which are relaxations events towards a
new such configuration. In avalanche dynamics, active (``fast'')
threshold values achieve improved stability only through the cooperative
behavior of neighboring sites. A critical
avalanche can only end through a long-range cooperation of many sites.
These avalanches, as in all extremal models of SOC, consist of a hierarchy of
sub-avalanches which are temporally and spatially contained within their
parent avalanches \cite{scaling}.  In Ref.~\cite{BoPa2} we showed how this
hierarchical structure in the analytically tractable multi-trait model
leads to the build-up of memory during an avalanche. There, we also
explored the ultrametric properties of the
self-organized critical state, and found that the
ultrametric separation between consecutive minima evolves to be
 power-law distributed. Thus, while imposed
explicitly to describe aging in glassy systems in Refs.~\cite{H+S},
hierarchical or ultrametric
structures emerge dynamically in some models of SOC and apparently lead
to similar aging behavior.

Our numerical results demonstrate that aging occurs not only in
glasses and other out of equilibrium systems slowly approaching their
asymptotic equilibrium state, but also occurs in some models of self-organized
 criticality.  Aside from
glasses or SOC, there exists a much wider range of important problems
exhibiting certain aspects of glassy behavior, for example the directed
polymer \cite{other}, protein folding,
learning, or optimization algorithms \cite{cnls-proc}.  The dynamics
of these problems is poorly understood and often conceptualized in
terms of a walk through a ``rugged landscape'' \cite{S+N}.  More than
adding another problem to that list, we believe that the avalanche
dynamics of some SOC models is far more accessible,  with many subtle
features considering their simplicity, that they might provide a basic
framework to study the emergence of slow relaxation and aging.  This
is certainly possible if the mechanism responsible for aging
is avalanche dynamics in systems approaching a critical state.

We acknowledge helpful correspondences with Mike Creutz.
SB acknowledges partial
support under DOE grant DE-FE02-95ER40923.

\end{multicols}
\end{document}